\documentclass[lettersize,journal]{IEEEtran}
\usepackage{amsmath,amsfonts}
\usepackage{algorithmic}
\usepackage{algorithm}
\usepackage{array}
\usepackage[caption=false,font=normalsize,labelfont=sf,textfont=sf]{subfig}
\usepackage{textcomp}
\usepackage{stfloats}
\usepackage{url}
\usepackage{verbatim}
\usepackage{graphicx}
\usepackage{cite}
\usepackage{xcolor}

\begin{document}
	
	\title{A new metric for the comparison of permittivity models in terahertz time-domain spectroscopy}
	
	\author{Melanie Lavancier, Nabil Vindas-Yassine, Juliette Vlieghe, Theo Hannotte, Jean-François Lampin, François Orieux, and Romain Peretti}
	
	
	
	\maketitle
	
	\begin{abstract}
		We present a robust method, as well as a new metric, for the comparison of permittivity models in terahertz time-domain spectroscopy (THz-TDS). \textcolor{black}{{In this work, we show that the signal is perturbed by a deterministic part that has to be either modeled or removed. Therefore, we perform an extensive perturbation/noise analysis of a THz-TDS system, we remove and model the unwanted deterministic perturbation/noises and implement them into our fitting process. This new method has now been implemented in our open-source software, Fit@TDS, available at : https://github.com/THzbiophotonics/Fit-TDS.}} \textcolor{black}{This work is the first step towards the derivation of uncertainties, and therefore the use of error bars. We hope that this will lead to performing analytical analysis with THz-TDS, as results obtained from different setups will be comparable. } Finally, we apply this protocol to the study of a $\alpha$-lactose monohydrate pellet in order to give more insight on the molecular dynamics behind the absorption peaks. The comparison with simulation results is made easier thanks to the probabilities derived from the metric. 
	\end{abstract}
	
	
	\section{Introduction}
	\IEEEPARstart{T}{Hz} THz spectroscopy of materials, ranging from $0.1$ to $10$ THz (\textit{i.e.} $3$ to $333$ cm$^{-1}$, \textit{i.e.} $0.1$ to $10$ ps), witnesses many chemicophysical processes in solids, liquids and gases \cite{Baxter2011,Eliet2021}. In solids, examples include dynamics of charge carrier in doped semiconductors \cite{Grischkowsky1990,Lampin2001,Houver2019} and 2D materials \cite{Gustafson2019}, probing the gap of superconductors \cite{Nuss1991,Brorson1996,Whitaker1994,Limaj2014}, and interrogating the phonon dynamics in molecular crystals \cite{Banks2023}. In solutions, molecular orientation \cite{Pedersen1992,Zalden2018}, the dynamics of the hydrogen bonds networks \cite{Schmidt2009} and their modification by solutes \cite{Nibali2014} are studied. In the gas phase \cite{Harde1994}, this range is even more selective than the fingerprint region \cite{Smith2015} and is therefore used to monitor chemical reactions \cite{Swearer2018}.

	Among THz spectroscopy techniques, terahertz time-domain spectroscopy (THz-TDS) is well-established and offers an extremely broad range and high dynamic range, making it widely used in laboratories whether it be with commercial or research setups. The measurements are made in the time-domain, which provides information on both the amplitude and the phase of the terahertz electric field, enabling the retrieval of the complex dielectric spectrum. To get further chemicophysical information, the experimental curves are fitted using permittivity models together with wave propagation in layers \cite{Tayvah2021}. The parameters of these models give quantitative insight into the motion and the state of the charges inside the sample, and on the signature of a compound.
	
	Today, there is no consensus on the fitting procedure to follow in order to retrieve those parameters. Most of the fits are done in the frequency domain using a least-square algorithm on the retrieved permittivity curves from experimental time-traces \cite{Ro/nne1997,Yada2008,Liebe1991,Moeller2009,Fitzgerald2014,Kindt1996}. However, THz-TDS outputs a signal as a function of time, which means that a Fourier transform has to be applied for this fitting method. In addition to complicated phase unwrapping, and to losing the phase for strong absorption \cite{Bernier2016}, this produces artifact-noises that can degrade the quality of the measurement and yield misleading results. In addition, while fitting in the frequency domain it is very difficult to  weight the function to minimize using the signal to noise ratio (or uncertainties) \cite{strutz2011}, therefore one cannot compare quantitatively two models for the same material or extract the uncertainty or error bars of the parameters of the model. The results of this work have the potential to unravel terahertz spectroscopy analytical full potential. In the future, our method could be used to improve the accuracy and precision of terahertz spectroscopy measurements, derive errors bars and to address some of the challenges remaining in the application of THz spectroscopy to biology and medicine \cite{Markelz2022}.
	
	\subsection*{Background description}
	
	Fit@TDS is an open source software based on time-domain fitting that has several advantages compared to other fitting methods \cite{Peretti2019}. It provides a more precise measurement of the thickness of the material, a higher precision on the refractive index retrieved, a better interpretation of the results and experiments, and a reliable and consistent retrieval of material parameters using permittivity models. The fitting process needs only two pieces of information:
	\begin{enumerate}
		\item A set of data containing the time-traces with and without sample ($E_{sample}(t)$ and $E_{ref}(t)$).
		\item A model depending on a set of parameters ${p_{i}}$ depicting how the sample transforms the reference pulse into the modeled one ($E_{model}\{p_i\}(t)$).
	\end{enumerate}

This modeled pulse is defined as the convolution between the transfer function of the experiment and the time trace of the reference:
\begin{equation}
	\label{eq1}
	E_{model}\{p_i\}(t) = T\{p_i\}(t)*E_{ref}(t)
\end{equation}
With $T\{p_i\}$ the transfer function of the experiment calculated with:
\begin{equation}
	\label{eq2}
	\tilde{T}\{p_i\}(\omega)=t_{a/s}t_{s/a}\frac{\exp{(-j\tilde{n}\{p_i\}\omega d/c)}}{1+r_{a/s}r_{s/a}\exp{(-2j\tilde{n}\{p_i\}\omega d/c)}}
\end{equation}

Where $t_{a/s}$ and $t_{s/a}$ correspond to the transmission coefficients ($a$ meaning "air" and $s$ meaning "sample"), $r_{a/s}$ and $r_{s/a}$ correspond to the reflection coefficients, $\tilde{n}\{p_i\}$ is the modeled complex refractive index of the sample and $d$ its thickness.

Then, an optimization algorithm is used in order to minimize the following error function:
\begin{equation}
	\label{eq3}
	\sum_{t=0}^{t=t_{max}}(E_{model}\{p_i\}(t)-E_{sample}(t))^2 dt
\end{equation}

The result given by this function changes according to the values that the model parameters take. The goal of the optimization algorithm is to find the parameters that minimize it using a least-square algorithm, which is widely used in THz-TDS data processing. However, as for other fitting methods, the permittivity model used for the fit is an ideal noiseless one. Indeed, the common approximation made is that the experimental noise is a Gaussian white noise, which won’t impact the fit.  \textcolor{black}Thus, when the signal includes any noise correlated to the signal, the algorithm mistakenly incorporates it into the analysis, potentially confusing it with genuine signal attributes. As a result, such algorithm struggles to distinguish between accurate and inaccurate models, as explained in  \cite{Mohtashemi2021, Dodge2020}. This becomes a problem when analyzing all sorts of samples. For example, the model chosen in the case of gases could change the shape of the absorption lines in the spectrum, leading to different interpretations as the broadening can be caused by collisions or by the Doppler effect. In liquids, and particularly for water, the question is even more important as there is no consensus on the correct model in the THz range. Several models are discussed, hence different physical interpretations, so it is necessary to quantitatively interpret the difference and better understand the pros and cons of each model.

Moreover, today, one of the main challenges of THz-TDS is to answer analytical problems \cite{Tayvah2021}. \textcolor{black}{{Yet, if the derivation of error bars for the refractive indices is provided in several software \cite{menlo}, the methodology for this derivation is not provided in the documentation as far as our knowledge extends. Therefore, such error bars only rarely appear in the published papers from the THz community on these topics. This prevents us from fully addressing analytical problems and therefore spreading the technique.}} This starts by the fact that, so far, the experimental noise is not correctly taken into account during the fitting procedure as raised by \cite{Mohtashemi2021}. We will start by presenting their main results that provide a method to overcome this issue, with the hypothesis that the noise has to be non-deterministic to be applied.

So, to take noise into account during the fitting process, we need to enter new information. When the input noise is convoluted by the transfer function (see equation \ref{eq1}), its contribution is found in all of the Fabry-Perot echoes in the sample time-trace. This is used by the
maximum likelihood estimator method explained in \cite{Mohtashemi2021}. For this method, there is an additional step of creating a noise correlation matrix that contains information about the noise both in the reference and sample time-traces including also information about the correlations. The creation of the first estimator for the noise correlation matrix is explained in the appendix section.

Now, it is possible to use a least-square algorithm to minimize an error function taking into account this noise correlation matrix (NCM) as all of the information on the noise is present. This new error function is \cite{Mohtashemi2021}:
\begin{equation}
	\label{eq4}
	Q(\mathbf{p_i})=[\mathbf{r}(\mathbf{p_i})]^T\mathbf{r}(\mathbf{p_i})
\end{equation}
Where
\begin{equation}
	\label{eq5}
	\mathbf{r}(\mathbf{p_i})=[\mathbf{M}_{noise}(\mathbf{p_i})]^{-\frac{1}{2}}[(\mathbf{E}_{model}(\mathbf{p_i})-\mathbf{E}_{sample})]
\end{equation}
and $\mathbf{M}_{noise}(\mathbf{p_i})$ is the noise correlation matrix, $\mathbf{E}_{model}(\mathbf{p_i})$ is the vector containing the fitted
time-trace data points, $\mathbf{E}_{sample}$ is the vector containing the experimental sample time-trace data points.

With this error function, the software is only fitting the data, not the artifact-noises. A further advantage to compare two models between one another is the ability to use the Akaike criterion \cite{Mohtashemi2021}:
\begin{equation}
	\label{eq6}
	AIC(\mathbf{p_i}) = Q(\mathbf{p_i})+2N_{p_i}
\end{equation}
where $N_{p_i}$ is the number of free parameters in the model. The smaller the result, the better the model. Moreover, this way, the optimal model is the one that has the best fit while not being over-parameterized. It is also useful to derive a probability from the criteria in order to have a quantitative comparison, where $\exp{((AIC_{min}-AIC_i)/2)}$ is proportional to the probability that the $i^{th}$ model minimizes the quantity of information. \textcolor{black}{{This equation is the last one from \cite{Mohtashemi2021}, and has certain limitations. Indeed, their method relies on a Monte Carlo study that cannot be applied on most experiments based on mechanical delay line setups. We would have to record a large number of accumulated time-traces consecutively (their example shows in their case of low time sample numbers 50 time-traces, but one needs to remember that it requires more time traces than the number of points in it for the covariance matrix to be invertible and therefore this number can easily exceed several thousands), which is impossible for stability reasons on most mechanical systems.}} Moreover, their transfer function is equal to $1$, meaning that they don’t have a sample inside their simulated experiment, they only compare two reference time-traces. Finally, they suppose that experimental noise is non-deterministic and artifact free, which is not true for real samples, as we will see in the rest of the article.

Hence, in this article, we are going to go further as we will adapt the method to a real THz-TDS setup and experiment and demonstrate that we now have a new metric to compare models by deriving their probability. It is important to notice that, to build the noise correlation matrix, the deterministic part of the noise has to be either modeled or removed. This means that we will start by doing an extensive noise analysis of our setup
to remove deterministic noises. Then, we will implement the new error function in our fit@TDS software as well as the Akaike criterion. Our errors will be quantified and can be compared not only within our own setup but also across different TDS setups. Finally, as a validation example, we will use the improved software to analyze a lactose sample. This is the "perfect" sample because of its use as a benchmark for THz-TDS setups, its availability and its complexity. Our goal is to give new capacities to the THz-TDS and to improve the quality of the retrieved information, such as being able to know if a spectral line is homogeneously or inhomogeneously broadened.

	\section{Most of the noise is deterministic}

\begin{figure}[!t]
	\centering
	\includegraphics[width=3.5in]{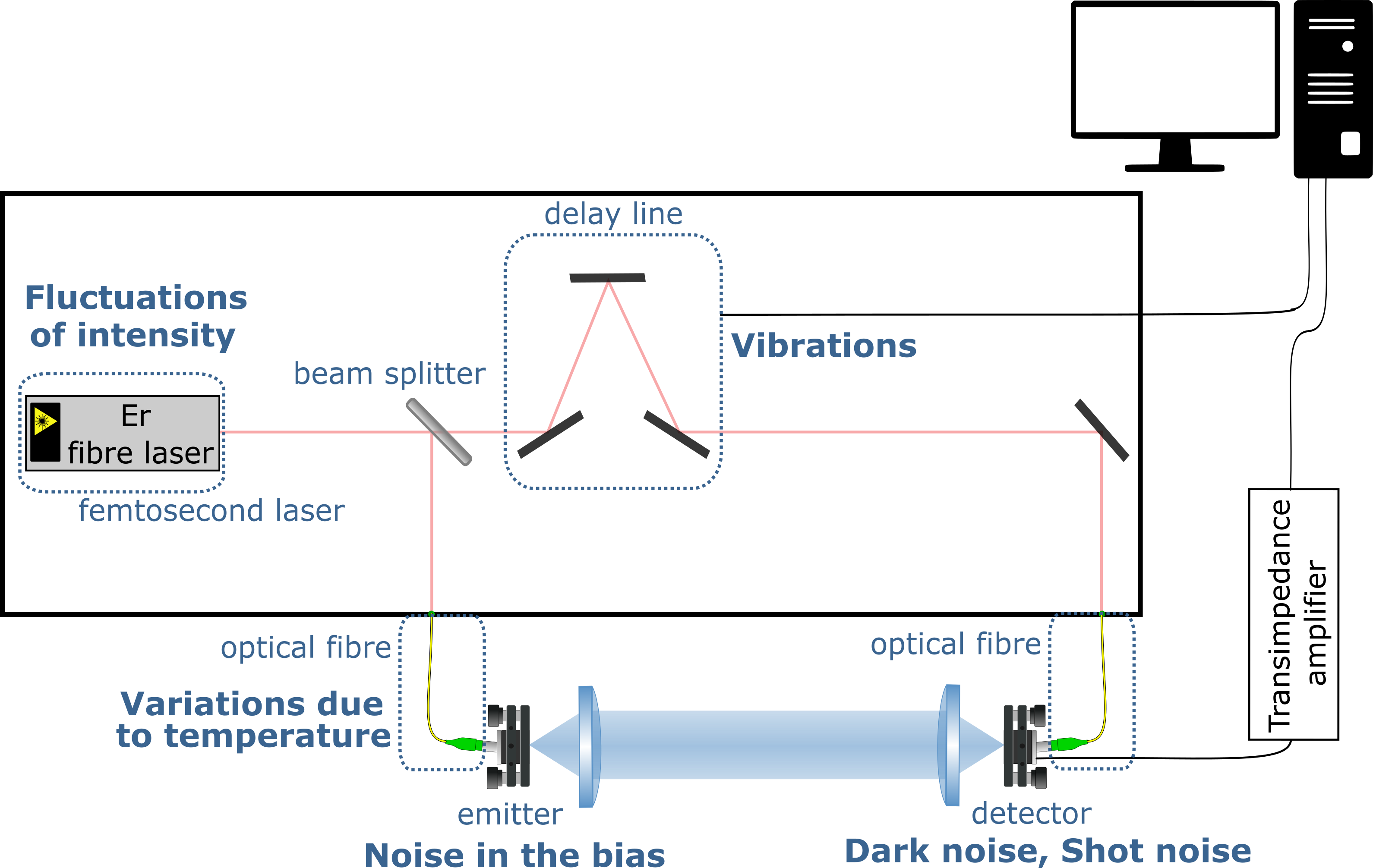}
	\caption{THz-TDS setup and its sources of experimental noise.}
	\label{fig1}
\end{figure}

We begin by clarifying the terminology used throughout the manuscript to describe various types of noise encountered in time-domain terahertz spectroscopy. In line with the general definition by Tuzlukov \cite{Tuzlukov2018}, we consider noise as any unwanted modification that the signal undergoes during capture, storage, transmission, processing, or conversion. Consequently, anything beyond the authentic signal itself falls under the classification of noise. Still, it's essential to recognize that not all noise sources are equivalent.
Among the different types of noise, we define:
\begin{itemize}
	\item \textbf{Artifact-noise: This category encompasses misleading disturbances originating directly from the experimental methodology employed. Specifically, in our experiments, artifact noise manifests as correlated feature, even in the absence of a terahertz signal at the detector.}
	\item \textbf{Deterministic noise: This noise, including the artifact-noise, displays consistent, repeatable behavior. When any noise has a non-zero average value, we refer to that component as the deterministic part of this noise.}
	\item \textbf{White noise: This ideal noise type exhibits a flat frequency spectrum up to a certain frequency exceeding the signal's highest frequency.}
	\item \textbf{Correlated noise (Non-white noise): This type deviates from the flat characteristic of white noise, indicating correlations between its frequency components.}
\end{itemize}

Experimental noise is the combination of different types of noises, such as dark noise or delay noise. The causes are numerous : fluctuations of the femtosecond laser intensity, vibrations in the delay line, noise in the bias of the photo-conductive antennas, reflections within components, noise in the trans-impedance amplifier, or influence of the temperature on the optical fibers as illustrated on figure \ref{fig1}. The sample can also be seen as a source of noise, or more generally non-reproducibility, as it may differ from the model due to its inhomogeneity, its scattering properties, or simply because of the temperature variations of its environment during the experiments. Uncorrelated Gaussian white noises can be minimized by increasing the number of averaging when data is collected but averaging during a long time can lead to other artifact-noises or drifts due to the variations in the environment so that a compromise has to be made. Moreover, other noises are correlated to the signal or determined by the signal and thus independent from the averaging and cannot be minimized by accumulation during the experiment but should instead be modeled.

There are many different contributing noises in our experiments. While some noises in our experiments may be uncorrelated and unpredictable, it's essential to analyze each type individually before assuming this applies to all. This approximation is valid for noises like thermal noise, Johnson noise, Shot noise, amplification noise, or laser noise because, regardless of their specific characteristics, the errors in THz-TDS experiments consistently exceed their individual contributions when appropriately averaged. However, other noises contribute more to the experimental noise and these are the ones we are going to analyze here to make sure that they are non-deterministic and that we will be able to apply the method from \cite{Mohtashemi2021}. It is important to keep in mind that the noise profiles are extremely dependent on the setup.

	\begin{figure}[!t]
	\centering
	\includegraphics[width=0.72\linewidth]{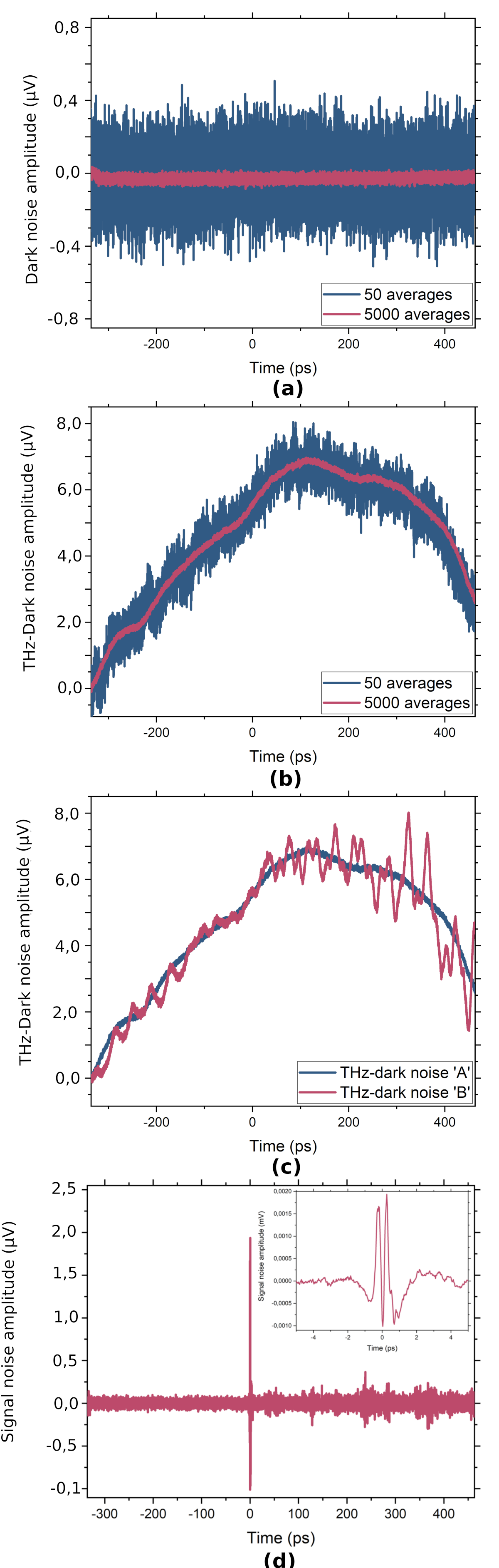}
	\caption{(a) Dark noise (no optical excitation on both photo-conductive antennas) recorded with two different averages, its contribution decreases when the averaging increases, independently of the time window. (b) THz-dark noise (no optical excitation on the emitter) recorded with two different averages, its contribution decreases when the averaging increases. However, it is a deterministic signal that depends on the chosen time window. (c) The two types of THz-dark noises, in blue without excitation on the emitter, in pink with optical excitation on the emitter and an absorber between the two antennas. (d) Remaining noise after removing the contributions of the THz-dark noise and fixing the delay. A zoom between $-5$ ps and $5$ ps is provided.}
	\label{fig2}
\end{figure}

\begin{itemize}
	\item{\textbf{Dark noise:} We call "dark noise" the random signal that is generated by the detector and amplifier. Even when no THz (or IR) pulse is shone onto the detector, some charge carriers exist in the semiconductor and hole-electron pairs can be created, generating a small current leading to this dark noise. This noise is not deterministic and its amplitude decreases as we increase the averaging number. Indeed, its noise spectral density goes from $5.10^{-6} \mu $V$\sqrt{Hz}$ for $50$ averages to $5.10^{-7} \mu $V$\sqrt{Hz}$ for $5000$ averages, which correspond to $4.10^{-5} \mu $V$\sqrt{Hz}*\sqrt{acc}$. As expected it does not depend on the time window scanned. In conclusion, it influences the fitting process, as it will be present in the Fabry-Perot echoes, but does not invalidate the fit methodology because of its amplitude and the fact that it is non-deterministic and uncorrelated. Therefore, it complies with the hypothesis.
	
	\item{\textbf{THz-dark noise:} \textcolor{black}By analogy with the definition of dark noise, we call ”THz-dark noise” the signal recorded by the THz-TDS in the absence of detection of the THz pulse. The THz-dark noise A corresponds to the laser being shined on the receiving antenna but not on the emitting antenna, and the THz-dark noise B to the laser being shined on both antennas but the THz beam is blocked by an absorber and cannot reach the receiver.} \textcolor{black}As we can see in figure \ref{fig2}.b, there are two components to this noise: a statistical noise varying when we replicate the experiment and an artifact-noise that is reproducible. The statistical noise is similar to the dark noise but with a higher amplitude. The artifact-noise depends on time (e.g. on the delay line position) and its amplitude contribution is most important at low frequencies, below 200 GHz, that is below the frequency limit of the system.} Hence, it is of utmost importance to remove this artifact-noise, as it does not go through the sample, otherwise it is easy to mistake it for real physical features when it is added to the pulse signal. A way to remove this noise would be to move the delay line to the emitter arm of the THz-TDS instead of the receiver one. However since we address the issue of noise for the broad use of THz-TDS we propose a way to deal with this noise without touching the system using signal processing. We implemented a high-pass filter into fit@TDS in order to get rid of the unwanted part of the data. This filter is a smooth step function, with the following formula:
	\begin{equation}
		\label{eq7}
		y[n]=0.5+0.5\tanh{(\frac{(f[n]-f_{cut})\alpha}{f_{cut}})}
\end{equation}
	Where $f$ is the frequency, $f_{cut}$ the cut-off frequency ($200$ GHz in our case), and $\alpha$ the sharpness of the filter, set at $10$ here. This can be adjusted to each setup according to the company’s claims considering the frequency range. With a cut-off frequency of
	$200$ GHz, the constant component of the noise is no longer there and the remaining THz-dark noise signal, have, after this filtering, a contribution of the same order of magnitude as the dark noise. In conclusion, this perturbation is no longer deterministic and thus complies with the hypothesis after the filtering.

\item{\textbf{Delay noise :} When recording the same time trace repeatedly, it appears that there is a small delay between each pulse of the order of a few femtoseconds, smaller than the pulse sampling time. We attributed it to a drift of the temperature in the optical fibers. Then, the measured delay corresponds to the average of the time drift during the time the spectrum was recorded. This small delay can lead to imprecision when retrieving the refractive index and/or the thickness of a sample. In order to lessen its contribution to the experimental noise, we implemented a feature in fit@TDS that allows the user to readjust two pulses so that there is no more delay between them. However, this feature should be used carefully to avoid overfitting. Typically, Fabry Perrot echoes should be present in the time trace to use it.}
\end{itemize}

The other sources of noise cannot be recorded separately. The only possibility is to find out what is left of the noise when all of the previous ones have been fixed. \textcolor{black}{{For this evaluation, we take $5$ consecutive measurements of a reference, that is with both antennas excited by the femtosecond laser and no absorber inside the setup. Then, we use the high pass filter with a $200$ GHz frequency cut-off and readjust the average delay between the pulses. If all of the noise contributions had been fixed besides the electric noise from the receiver and amplifier, the difference between one of these 5 measurements and their mean would give a Gaussian white noise. However, figure \ref{fig2}.d shows that there is still a small noise left, which is not deterministic but shows an amplitude correlated to the signal.}}

\begin{figure}[!t]
	\centering
	\includegraphics[width=3.5in]{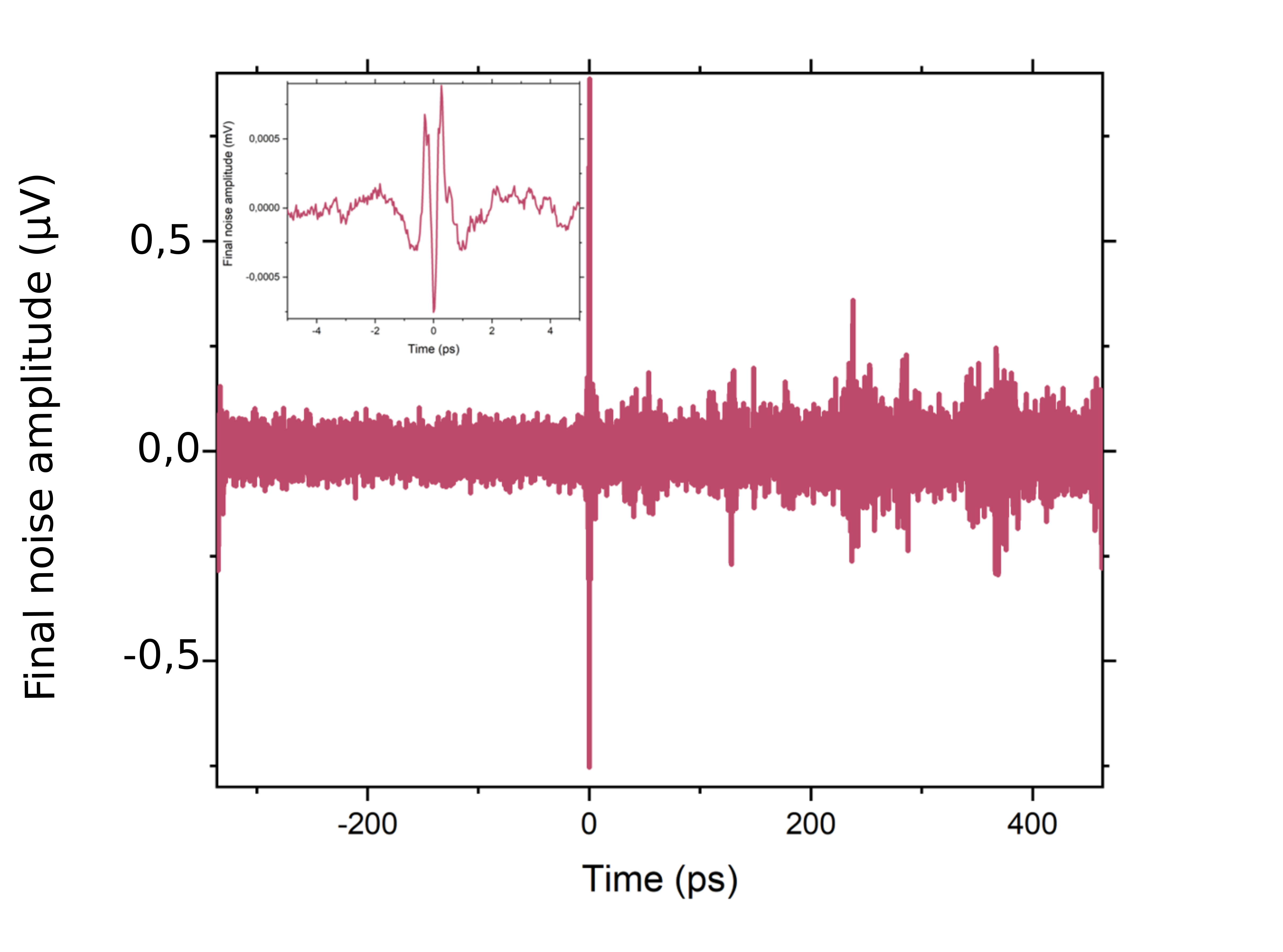}
	\caption{Experimental noise recorded on our Menlosystems THz-TDS setup after having removed the contribution from other sources of noise \textcolor{black}{{and having fitted the leftover noise with a phenomenological model.}} A zoom between $-5$ ps and $5$ ps is also provided.}
	\label{fig3}
\end{figure}

\textcolor{black}{{This noise corresponds to a sum of contributions coming from the other sources of noise in the setup, such as delay drift (we only adjusted the mean time drift), shot noise and laser power fluctuations. This combination of noises has a mean value depending on time and an amplitude correlated to the signal as seen on figure \ref{fig3}.}} It means that noise at one particular time contains information about the noise at neighboring times. We use a phenomenological model to fit this leftover noise in order to remove its influence on the data.

Once this final step has been done, we can see on figure \ref{fig3}, that most of the noise has been eliminated. There is only a small contribution left. We found that this signal corresponds to the flickered noise, which is non-deterministic. To use the method proposed by \cite{Mohtashemi2021}, based on a Monte Carlo algorithm, ones needs to know the evolution of noise variance as a function of time. This corresponds to the envelope of this final noise. Therefore, the corresponding noise correlation matrice can be retrieved to use the new error function during the fitting process. We insist on the fact that this study allowed us to remove unwanted correlated noise, but it is a first approach. Indeed, to stay close to the experiment and avoid other artifact-noises due to long measurement times, we only took five measurements, and there is also improvement to be done on our acquisition software that does not allow to record the spectra one by one at this moment (which complicates the correction of delay drifts). Moreover, we will work in the future on a cleaner modeling of these noises, namely on the leftover noise fitted with a phenomenological model. Here, our main ambition is to clean the THz-TDS data in order to improve the extraction of model parameters.
	
	\section{The shapes of the lactose absorption peaks}
	The monohydrate-crystallized form of lactose, called $\alpha$-lactose monohydrate, is widely used in THz-TDS thanks to its easily recognizable absorption peaks at $0.53$, $1.19$ and $1.37$ THz. Moreover, due to its availability and low cost, lactose samples are a
	"textbook case", and often used to test THz spectroscopy equipment \cite{peiponen2012terahertz}, serve as a "first-try" sample in minor-volume detection techniques \cite{Peretti2019,mitryukovskiy2019shining}, and are used as a mixture compound in approving methods for content quantification \cite{qiao2014identification}. Despite the abundance of literature on the subject, lactose remains a challenging material for analysis due to its polymorphic nature, which complicates accurate simulation of this molecular crystal. Theoretical explanations of lactose's terahertz (THz) spectrum are notably scarce, relying primarily on gas-phase Density Functional Theory (DFT) calculations that focus on individual lactose molecules rather than crystalline structures, thus overlooking collective vibrations. Recent investigations \cite{Mitryukovskiy2021} have contributed to this area by comparing DFT simulations of crystal structures across various lactose polymorphs, aligning them with experimental data using boundary periodic conditions \cite{banks2021necessity}. However, an overlooked aspect is that lactose samples, often in compressed powder form, introduce defects and a degree of disorder that warrants further study.
	
	\subsection*{Experimental section}
	
\begin{figure}[!t]
	\centering
	\includegraphics[width=0.75\linewidth]{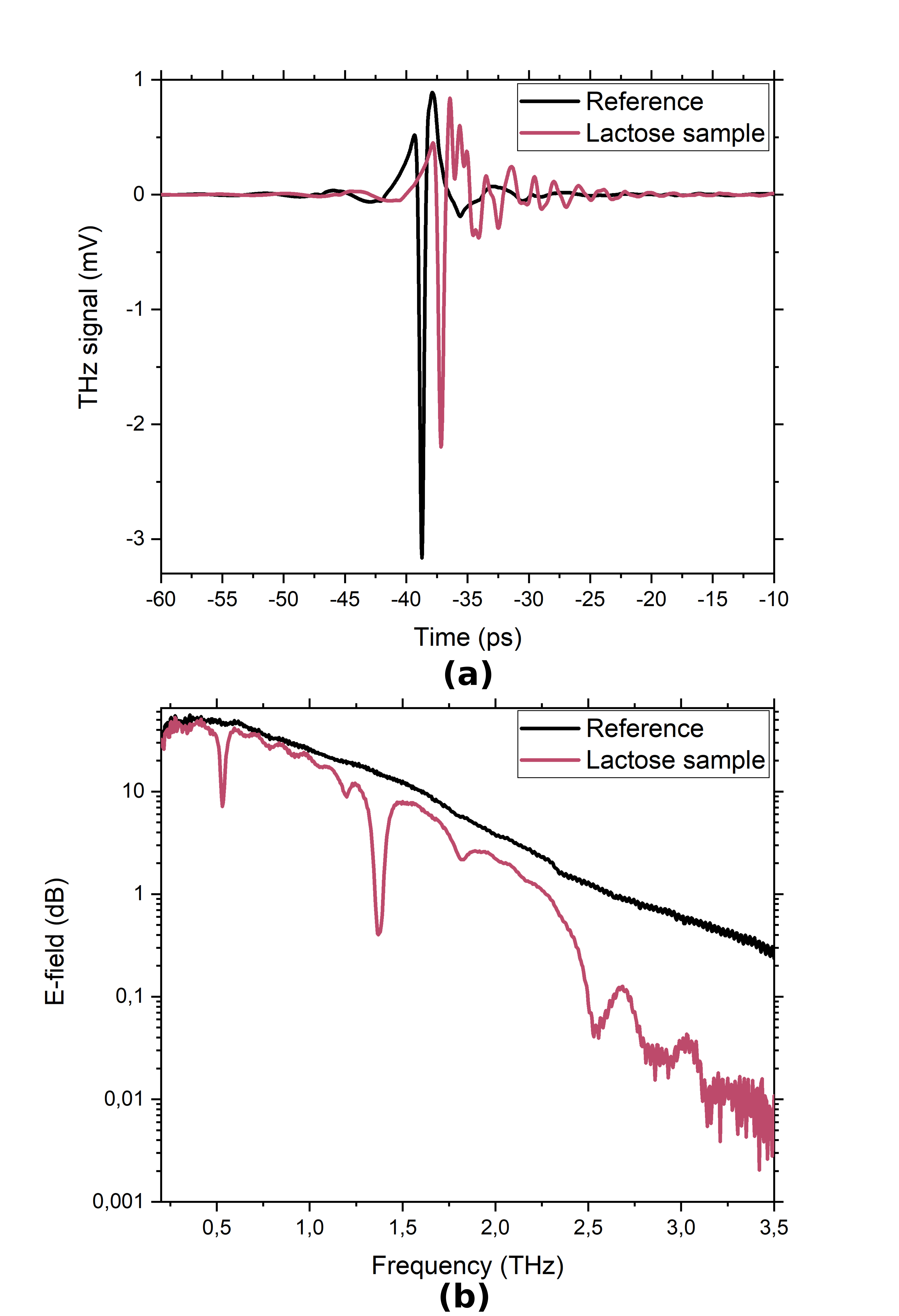}%
	\caption{Time-trace and spectrum of the reference (\textit{i.e.} the THz pulse goes through the holder without pellet inside) and the sample. The time-trace measurement was actually made on a larger time range (between $-80$ and $120$ ps but it is zoomed in for visibility.}
	\label{fig4}
\end{figure}
	
	The lactose powder ($\alpha$-lactose monohydrate) was purchased from Sigma-Aldrich Co. Ltd. ($\ge 99.9\%$ total $\alpha$-lactose basis including less than $4\%$ $\beta$-lactose) and was pressed into a pellet (there was no dilution) with a $13$mm diameter and $600$ $\mu$m thickness. The measurement was made with a commercial Terasmart setup by Menlosystems, shown in figure \ref{fig1}, placed inside a nitrogen-purged box. The laser frequency repetition is 100MHz. To the two collimating lenses with a 50mm focal length we added two focusing lenses with the same focal length since the diameter of the pellet is smaller than the diameter of the THz collimated beam. The sample was placed in the middle of the setup, held inside a home-made metallic holder. The reference was taken with the same holder without any pellet inside. In order to build the noise correlation matrices, we took two consecutive
	measurements in both cases. A time trace of the reference and one of the sample and their corresponding spectra are shown in figure \ref{fig4}. The oscillations we see on the lactose time trace are due to the Fabry-Perot echoes inside the pellet as well as the absorption, visible under
	the form of a peak on the spectrum.
	
	\subsection*{The fitting process}
	
	The first step after having acquired the data is to do the pre-processing explained in the previous section. Since we have two consecutive measurements of the reference and two consecutive measurements of the sample, we infer the envelope of the remaining noise for each set by subtracting the two and applying the high-pass filter, correcting the delay and fixing the noise proportional to the signal and its second derivative. From the two signals left (one for the reference and one for the sample), we build a noise correlation matrix (NCM) following the protocol explained in \cite{Lavancier2021} and in the appendix section. This NCM will then be used as input in fit@TDS during the fitting process \textcolor{black}{{using a dual annealing algorithm}}.
	
	In the literature, the lactose absorption peaks are usually modeled by a single Lorentz oscillator. However, since there are different dynamics behind each peak, there is no reason why they should all have the same shape. \textcolor{black}{{Therefore, our goal is to use our new metric to compare, for each peak, the Lorentz model : 
	\begin{equation}
		\label{eq8}
		\chi_{Lorentz}(\omega)=\frac{\Delta\epsilon\omega_0^2}{\omega_0^2-\omega^2+j\omega\gamma}
	\end{equation}
	To the Voigt model :
		\begin{align}
			\label{eq9}
			\chi_{Voigt}(\omega) &= \frac{1}{\sqrt{2\pi}\sigma} \int_{-\infty}^{\infty} \frac{\exp{(\frac{-(\omega-\omega')^2}{2\sigma^2})}\Delta\epsilon\omega_0^2}{\omega_0'^2-\omega'^2+j\omega'\gamma}d\omega' \\
			&= \chi_{Lorentz}(\omega)*\frac{1}{\sqrt{2\pi}\sigma}\exp{(\frac{-\omega^2}{2\sigma^2})}
	\end{align}
Where $\Delta\epsilon$ is the strength of the oscillator, $\omega_0$ is the center frequency of the Lorentz oscillator, $\gamma$ is the width of the Lorentz oscillator and $\sigma$ is the width of the Gaussian broadening.
	This will give us more information on the microscopic structure of the sample. Indeed, an ideal crystal would have Lorentz absorption peaks, but defects inside the structure broaden the peaks, creating the need for a Voigt model.}}
	
We fitted our data peak by peak, from the ones that have the most energy to the ones that have less energy. \textcolor{black}{{First, we paid attention to the fact that each feature we fitted sequentially did not spread towards any other feature that had not yet been included, which could perturb the fit at the next step. In addition, we let a small freedom to all of the parameters of the fit (locally in the parameter space) to adapt to any new added feature. For the shape of the peak, we tested all of the models and chose the one that was most probable according to the Akaike criterion.}} Indeed, if the value of the criterion alone does not give much indication, it allows us to compare models : a lower Akaike criterion means a better suited model. Moreover, it also includes a penalty term that is an increasing function of the number of estimated parameters. The penalty discourages overfitting, which is desired because increasing the number of parameters in the model almost always improves the goodness of the fit. The first step of the fitting process is to run a fit without any model in order to have a base criterion. If the fit was perfect, the Akaike criterion would be equal to the number of points (in our case $6.0e+03$). When no model is entered (only a constant value for the permittivity and a value for the thickness of the pellet), the fit is far from perfect and the value of the criterion is $6.1e+08$. Now, our goal is to reduce this criterion and, more importantly, to find the model that reduces it the most. We found that there was a need to fit the broad high frequency losses first. \textcolor{black}{{Thus, we decided to fit it phenomenologically with a continuum model :
\begin{multline}
	\label{eq10}
	\chi(\nu) = \chi_{Lorentz}(\nu)* \\
	[1-(1-\frac{\nu}{\nu_1}-\frac{nu^2}{2\nu_1^2}).\exp(-\frac{\nu}{\nu_1})]. \\
	H(\nu-\nu_0,1)
\end{multline}
Where $\nu$ is the frequency, $\nu_1=\frac{k}{h}$, $h$ is Planck constant, $k$ is a continuum constant, and $H$ is the Heaviside function, meaning that $H(\nu-\nu_0,1)=1$ when $\nu \ge \nu_0$.
This model reduced the Akaike criterion to $2.5e+08$. It compensates for all absorption losses since no peaks have been added yet, but its parameters will adjust as the fitting process progresses.}}

\textcolor{black}{{Now, we can focus on the two main absorption peaks of lactose. In the appendix, we provide a table that showcases the results for the peaks that absorb the most, which are located at $0.53$ and $1.37$ THz. In all cases, the criterion has been reduced significantly, which proves the need to fit these two absorption peaks. The configurations that possess the lowest Akaike results, which means that they are the best suited models (between the ones that we have tested), are the ones that use a Voigt model for the first peak and either a Lorentz or Voigt model for the second one. We ended up choosing the Lorentz model for the second peak because of the fact that the width of the Gaussian in the Voigt model is really small ($0.1$ GHz) thus negligible considering the frequency resolution ($5$ GHz). The parameters of the continuum are now: $\Delta\epsilon = 0.588$, $\omega_0=3.389$ THz, $\gamma=2.207$ THz.}}

\textcolor{black}{{For pedagogic reasons we discussed thoroughly the first four steps of the fitting process while the remaining steps are quickly discussed for the sake of concision. These next steps are to add absorption peaks one by one until the criterion does not decrease anymore.}}

\subsection{The final fit}

To continue taking the experimental noise into account, it is important to keep in mind that the delay noise and the leftover noise still have to be modelled. During the pre-processing of the data, we only fixed them for the two consecutive measurements of the reference, and then of the sample. What we correct is the average value of the drift between two measurements, which means that if we look at two different measurements there is no reason why this value should be the same. As a result, we have to incorporate it in our fit as well. However, if we add these parameters too soon in the fitting process, they tend to compensate other losses. Hence, it is important to follow the magnitude of the Akaike criterion. Each feature should be added in decreasing magnitude order.

After having fitted the two main absorption peaks as well as the high frequency absorption, according to the Akaike criterion, the noise terms can be added to the fitting process. Finally, due to theoretical considerations \cite{Mitryukovskiy2021}, we decided to replace the Lorentz model for the $1.37$ THz absorption peak by a doublet (which was observed in DFT simulations), as well as the $1.8$ THz peak and fitted them. \textcolor{black}{{The probability derived from the Akaike criteria for models with and without the doublet confirm that the model with doublet is the most probable.}}

\begin{table*}[!t]
	\caption{The optimized model parameters of lactose. Other retrieved information is: $\epsilon_\infty = 2.739$ ; thickness of the pellet is $599.5 \mu m$ (coherent with the measured one); Average delay is $-17.5$ fs ; Coefficients of the noise proportional to the signal and its derivatives are $a = 3.393e-05$ and $b = 6.944e-04$ ; Final Akaike criterion is $2.028e+06$. \label{table2}}
	\centering
	\begin{tabular}{c c c c c c}
		Feature & Model & $\Delta\epsilon$ & $\omega_0$ & $\gamma$ & $\sigma$ \\
		& & & (THz) & (GHz) & (GHz)\\
		\hline
		$1^{st}$ peak & Voigt & $0.046$ & $0.531$ & $23.17$ & $4.280$ \\
		\hline
		$3^{rd}$ peak & Voigt & $3.822e-03$ & $1.195$ & $43.88$ & $0.837$ \\
		\hline
		$2^{nd}$ peak & Doublet(Voigt) & $3.839e-03$ & $1.264$ & $316.0$ & $3.670$ \\
		& & $0.029$ & $1.370$ & $46.56$ & $10.47$ \\
		\hline
		$4^{th}$ peak & Doublet (Voigt) & $9.764e-06$ & $1.767$ & $1.138e-03$ & $695.5$ \\
		 & & $5.379e-03$ & $1.818$ & $104.97$ & $6.362$ \\
		\hline
		$5^{th}$ peak & Doublet(Voigt) & $5.962e-03$ & $2.527$ & $10.68$ & $49.28$ \\
		 & & $0.067$ & $2.594$ & $51.05e-07$ & $1.417e+03$ \\
		\hline
		$6^{th}$ peak & Voigt & $0.085$ & $2.901$ & $0.274$ & $889.2$ \\
		\hline
		$7^{th}$ peak & Voigt & $0.119$ & $3.409$ & $44.04$ & $475.7$ \\
		\hline
		High frequency absorption & Continuum & $0.184$ & $0.258$ & $5.460$ & \\
		\hline
	\end{tabular}
\end{table*}

The final model parameters are given in table \ref{table2}. The final Akaike criterion achieved is $2.0e+06$, which is strongly lower than the base criterion. However, we have not yet reached a "perfect" fit, and it could certainly be improved by a better modeling of the noise and a more accurate model for the high frequency losses, \textcolor{black}{{or a more sophisticated model for the vibration of the molecular crystal lattice.}} Moreover, we do not take into account the scattering or the fact that the incident beam is not perfectly perpendicular to the studied sample. Finally, the fabrication of the pellet itself has an impact on the measured data and thus the model, because of the metamictisation due to the applied pressure for instance. Therefore, both the modeling and the experiment could be improved to lower the Akaike criterion. Still, the results obtained match the data available in the literature \cite{Federici2005, Taday2003}. Indeed, \cite{Federici2005} find that the peaks are located at $0.54$THz, $1.20$THz, $1.38$THz, $1.82$THz, $2.54$THz, $2.87$THz, and $3.29$THz.

\begin{figure}[!t]
	\centering
	\includegraphics[width=0.75\linewidth]{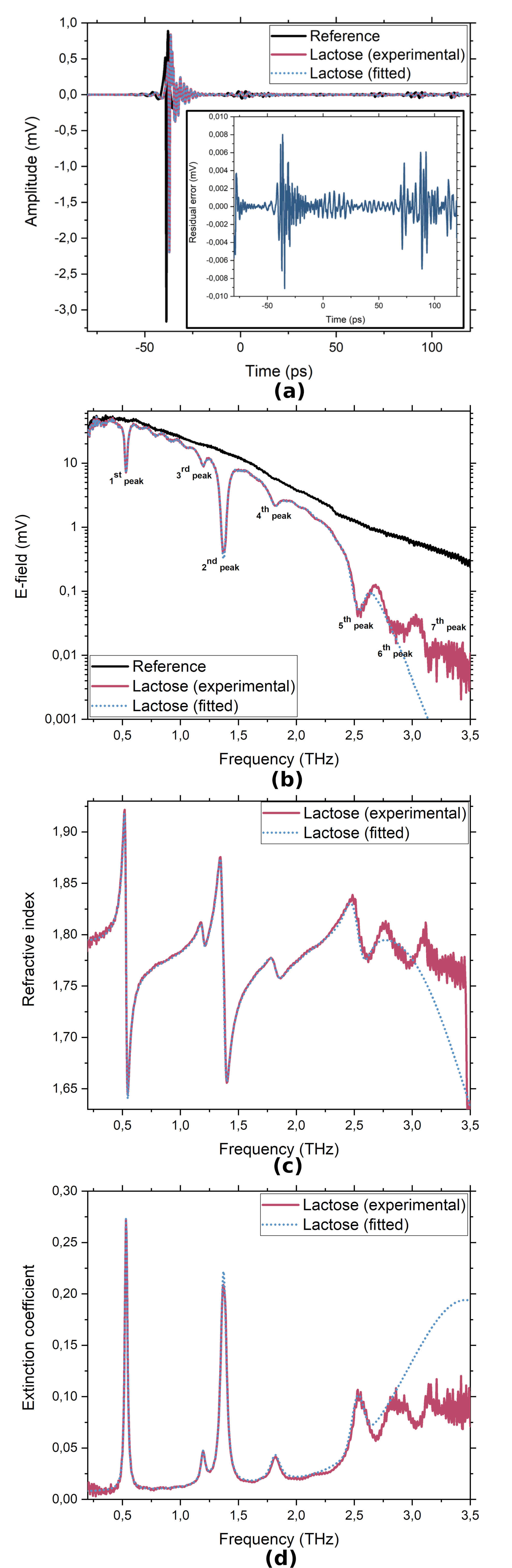}
	\caption{Results of the fitting process, illustrating the parameters retrieved in table \ref{table2}. A) Reference (in black), experimental lactose sample (in pink), and fitted lactose sample (in light blue dots) time traces. In the inset is the residual error of the fit. B) :Reference (in black), experimental lactose sample (in pink), and fitted lactose sample (in blue dots) spectra. The number of the absorption peaks correspond to the parameters given in table \ref{table2}. C) Refractive index retrieved from experimental data and from the fitted data. D) Extinction coefficient retrieved from experimental data and from the fitted data.
}
		\label{fig5}
	\end{figure}

The fitted spectrum, as well as the real and imaginary parts of the refractive index are shown in figure \ref{fig5}. Above $2.6$ THz, the fitted curve strays from the experimental data, which is expected since they become noisier. Moreover, the absorption peaks are wider, which can mean that the Lorentz and Voigt model do not describe the dynamics correctly any more and another model may be more appropriate. With more input from simulations we would be able to adjust this part of the fit.

The primary objective of this study was to demonstrate the capacity of our method to extract physical parameters using models capable of accommodating variations in shape. Consequently, we chose to assess the models individually, refraining from conducting direct statistical comparisons between them. It is worth noting that our current research represents an initial phase in the development of our model, and the distinctions drawn within this paper should not be regarded as definitive. In future investigations, we anticipate conducting direct statistical comparisons between the models to ascertain whether the observed disparities hold statistical significance, thereby confirming that these differences are not merely the result of random chance.

\section{Conclusion}
Time-domain fitting, in addition to being closer to the experiment, makes the noise evaluation and standard deviation evaluation easier and therefore a more practical way to use the reliable method proposed in \cite{Mohtashemi2021}. However, to analyze data even more rigorously in THz-TDS, taking non-Gaussian noise and artifact-noises into account during the fitting process is of utmost importance. Hence, we have to have an extensive knowledge of the noise arising from the setup in order to build noise correlation matrices. By taking into account this additional information in the pre-processing of the data as well as during the fitting process, we ensure that the retrieved permittivity models describe accurately the analyzed sample. Moreover, by introducing the Akaike criterion, we are now able to derive the relative probability of the models accuracy and
therefore it is possible to compare them quantitatively. Furthermore, we are able to use this metric to perform a better fit a set of data, or even to compare results from different setups.

Still, this first approach on the noise analysis of a THz system raises new questions, especially on the presence and origin of the leftover noise. It is now fitted with a phenomenological model and has to be investigated further. In the future, more measurements should be taken using an automated process independent from any averaging done by the constructors’ software. Moreover, thanks to the implementation of the covariance matrix
and the precision matrix, more elaborate noise estimators will be implemented.

Here, the application of this first method to the analysis of a $\alpha$-lactose monohydrate pellet gave insight on the molecular dynamics behind the absorption peaks. The comparison with simulation results, namely regarding the presence of the doublet, is also made easier thanks to the probabilities derived from the Akaike criterion. This shows the usefulness of this new metric and encourages us to use the same method for other complex samples. We hope that this study, in the spirit of several other recent ones \cite{Tayvah2021,Lee2023}, is one of the first steps to give new capacities to THz-TDS, and will help improve the quality of the retrieved information as it will help in setting the error bars. Finally, we provide the open-source fit@TDS software at : https://github.com/THzbiophotonics/Fit-TDS.

	{\appendix[Creating the noise convolution matrices]
		After the pre-processing of the data, since we have two consecutive measurements for the reference and two consecutive
		measurements of the sample, we retrieve the remaining noise for each set by subtracting the two and applying the high-pass filter, correcting the delay and fitting with the phenomenological model.
		
		The next step consists in retrieving the envelopes of the remaining signals (from the reference and sample). The part left inside these envelopes is a Gaussian noise, thus all the information is contained in the envelope. Then, we take the square of this envelope. Finally, we created a Matlab program with the following steps:
		\begin{enumerate}
			\item With the two time-traces (from the reference and the sample), we start by calculating the experimental transfer function of the experiment : $TF(\omega)=\frac{E_{lactose}(\omega)}{E_{ref}(\omega)]}$
			\item We begin a Monte Carlo study, meaning that we repeat the following steps a large number of times:
			\begin{itemize}
				\item We multiply each noise envelope by a random vector, to obtain $\sigma_{noise,ref}(t)$ and $\sigma_{noise,sample}(t)$.
				\item For the reference, we apply the transfer function to this result in the frequency domain, and then go back to time domain: $\sigma_{noise,TFref}(t)=FT^{-1}[FT[\sigma_{noise,ref}(t)].TF(\omega)]$ (where $FT$ is short for Fourier Transform). This indicates the contribution of the
				reference noise in the Fabry Perot echoes as well as in the principal pulse.
				\item We multiply each result by its transpose in order to obtain a matrix: $U_{ref}=\sigma_{noise,TFref}(t).\sigma_{noise,TFref}^{T}(t)$ and $U_{sample}=\sigma_{noise,sample}(t).\sigma_{noise,sample}^{T}(t)$
			\end{itemize}
			\item Once the Monte Carlo study over, we add all the results to obtain the final matrices and finally add the reference and sample noise convolution matrices: $U_{tot}=U_{ref}+U_{sample}$.
			\item We invert and take the square-root of this matrix to obtain the final NCM: $U_{final}=[U_{tot}]^{-\frac{1}{2}}$.
	\end{enumerate}}

{\appendix[The first steps of the fitting process]
	
	Table \ref{table1} shows the results obtained when we try different models for the first two peaks of the lactose sample. 
	
	\begin{table*}[!t]
		\caption{The optimized parameters for the possible models for the first two absorption peaks of lactose. \label{table1}}
		\centering
		\begin{tabular}{c c c c c c c c}
			Model & $\epsilon_\infty$ & $\Delta\epsilon$ & $\omega_0$ & $\gamma$ & $\sigma$ & Akaike criterion & Relative probability \\
			& & & (THz) & (GHz) & (GHz) & & \\
			\hline
			$1^{st}$ peak: Lorentz & $2.614$ & $0.050$ & $0.530$ & $25.20$ & & $6.3e+06$ & $10^{-2.0e4}$ \\
			$2^{nd}$ peak: Lorentz & & $0.029$ & $1.370$ & $47.52$ & & & \\
			\hline
			$1^{st}$ peak: Lorentz & $2.616$ & $0.050$ & $0.530$ & $25.20$ & & $6.3e+06$ & $10^{-2.0e4}$ \\
			$2^{nd}$ peak: Voigt & & $0.029$ & $1.370$ & $47.50$ & $9.246$ & & \\
			\hline
			$1^{st}$ peak: Voigt & $2.611$ & $0.049$ & $0.530$ & $23.32$ & $3.820$ & $6.2e+06$ & $1$ \\
			$2^{nd}$ peak: Lorentz & & $0.029$ & $1.370$ & $47.36$ & & & \\
			\hline
			$1^{st}$ peak: Voigt & $2.611$ & $0.049$ & $0.530$ & $23.32$ & $3.820$ & $6.2e+06$ & $10^{-0.5}$ \\
			$2^{nd}$ peak: Voigt & & $0.029$ & $1.370$ & $47.36$ & $0.1$ & & \\
			\hline
		\end{tabular}
\end{table*}}


\bibliographystyle{IEEEtran}

\bibliography{test.bib}

\begin{thebibliography}{10}
\providecommand{\url}[1]{#1}
\csname url@samestyle\endcsname
\providecommand{\newblock}{\relax}
\providecommand{\bibinfo}[2]{#2}
\providecommand{\BIBentrySTDinterwordspacing}{\spaceskip=0pt\relax}
\providecommand{\BIBentryALTinterwordstretchfactor}{4}
\providecommand{\BIBentryALTinterwordspacing}{\spaceskip=\fontdimen2\font plus
\BIBentryALTinterwordstretchfactor\fontdimen3\font minus
  \fontdimen4\font\relax}
\providecommand{\BIBforeignlanguage}[2]{{%
\expandafter\ifx\csname l@#1\endcsname\relax
\typeout{** WARNING: IEEEtran.bst: No hyphenation pattern has been}%
\typeout{** loaded for the language `#1'. Using the pattern for}%
\typeout{** the default language instead.}%
\else
\language=\csname l@#1\endcsname
\fi
#2}}
\providecommand{\BIBdecl}{\relax}
\BIBdecl

\bibitem{Baxter2011}
J.~B. Baxter and G.~W. Guglietta, ``Terahertz spectroscopy,'' \emph{Analytical
  chemistry}, vol.~83, no.~12, pp. 4342--4368, 2011.

\bibitem{Eliet2021}
S.~Eliet, A.~Cuisset, F.~Hindle, J.-F. Lampin, and R.~Peretti, ``Broadband
  super-resolution terahertz time-domain spectroscopy applied to gas
  analysis,'' \emph{IEEE Transactions on Terahertz Science and Technology},
  vol.~12, no.~1, pp. 75--80, Jan. 2021.

\bibitem{Grischkowsky1990}
D.~Grischkowsky, S.~Keiding, M.~van Exter, and C.~Fattinger, ``Far-infrared
  time-domain spectroscopy with terahertz beams of dielectrics and
  semiconductors,'' \emph{Journal of the Optical Society of America B}, vol.~7,
  no.~10, p. 2006, Oct. 1990.

\bibitem{Lampin2001}
J.~F. Lampin, L.~Desplanque, and F.~Mollot, ``Detection of picosecond
  electrical pulses using the intrinsic franz{\textendash}keldysh effect,''
  \emph{Applied Physics Letters}, vol.~78, no.~26, pp. 4103--4105, Jun. 2001.

\bibitem{Houver2019}
S.~Houver, L.~Huber, M.~Savoini, E.~Abreu, and S.~L. Johnson, ``2d {THz}
  spectroscopic investigation of ballistic conduction-band electron dynamics in
  {InSb},'' \emph{Optics Express}, vol.~27, no.~8, p. 10854, Apr. 2019.

\bibitem{Gustafson2019}
J.~K. Gustafson, P.~D. Cunningham, K.~M. McCreary, B.~T. Jonker, and L.~M.
  Hayden, ``Ultrafast carrier dynamics of monolayer ws2 via broad-band
  time-resolved terahertz spectroscopy,'' \emph{The Journal of Physical
  Chemistry C}, vol. 123, no.~50, pp. 30\,676--30\,683, Nov. 2019.

\bibitem{Nuss1991}
M.~C. Nuss, P.~M. Mankiewich, M.~L. O'Malley, E.~H. Westerwick, and P.~B.
  Littlewood, ``Dynamic conductivity and coherence peak
  {inYBa}2cu3o7superconductors,'' \emph{Physical Review Letters}, vol.~66,
  no.~25, pp. 3305--3308, Jun. 1991.

\bibitem{Brorson1996}
S.~D. Brorson, R.~Buhleier, I.~E. Trofimov, J.~O. White, C.~Ludwig, F.~F.
  Balakirev, H.-U. Habermeier, and J.~Kuhl, ``Electrodynamics of
  high-temperature superconductors investigated with coherent terahertz pulse
  spectroscopy,'' \emph{Journal of the Optical Society of America B}, vol.~13,
  no.~9, p. 1979, Sep. 1996.

\bibitem{Whitaker1994}
Y.~L. John F.~Whitaker, Feng~Gao, ``Terahertz bandwidth pulses for coherent
  time-domain spectroscopy,'' \emph{Proceedings Volume 2145, Nonlinear Optics
  for High-Speed Electronics and Optical Frequency Conversion}, 1994.

\bibitem{Limaj2014}
O.~Limaj, F.~Giorgianni, A.~D. Gaspare, V.~Giliberti, G.~de~Marzi, P.~Roy,
  M.~Ortolani, X.~Xi, D.~Cunnane, and S.~Lupi, ``Superconductivity-induced
  transparency in terahertz metamaterials,'' \emph{{ACS} Photonics}, vol.~1,
  no.~7, pp. 570--575, Jun. 2014.

\bibitem{Banks2023}
P.~A. Banks, E.~M. Kleist, and M.~T. Ruggiero, ``Investigating the function and
  design of molecular materials through terahertz vibrational spectroscopy,''
  \emph{Nature Reviews Chemistry}, pp. 1--16, 2023.

\bibitem{Pedersen1992}
J.~Pedersen and S.~Keiding, ``{THz} time-domain spectroscopy of nonpolar
  liquids,'' \emph{{IEEE} Journal of Quantum Electronics}, vol.~28, no.~10, pp.
  2518--2522, 1992.

\bibitem{Zalden2018}
P.~Zalden, L.~Song, X.~Wu, H.~Huang, F.~Ahr, O.~D. Mücke, J.~Reichert,
  M.~Thorwart, P.~K. Mishra, R.~Welsch, R.~Santra, F.~X. Kärtner, and
  C.~Bressler, ``Molecular polarizability anisotropy of liquid water revealed
  by terahertz-induced transient orientation,'' \emph{Nature Communications},
  vol.~9, no.~1, May 2018.

\bibitem{Schmidt2009}
D.~A. Schmidt, Özgür Birer, S.~Funkner, B.~P. Born, R.~Gnanasekaran, G.~W.
  Schwaab, D.~M. Leitner, and M.~Havenith, ``Rattling in the cage: Ions as
  probes of sub-picosecond water network dynamics,'' \emph{Journal of the
  American Chemical Society}, vol. 131, no.~51, pp. 18\,512--18\,517, Nov.
  2009.

\bibitem{Nibali2014}
V.~Conti~Nibali and M.~Havenith, ``New insights into the role of water in
  biological function: Studying solvated biomolecules using terahertz
  absorption spectroscopy in conjunction with molecular dynamics simulations,''
  \emph{Journal of the American Chemical Society}, vol. 136, no.~37, pp.
  12\,800--12\,807, Sep. 2014.

\bibitem{Harde1994}
H.~Harde, N.~Katzenellenbogen, and D.~Grischkowsky, ``Terahertz coherent
  transients from methyl chloride vapor,'' \emph{Journal of the Optical Society
  of America B}, vol.~11, no.~6, p. 1018, Jun. 1994.

\bibitem{Smith2015}
R.~M. Smith and M.~A. Arnold, ``Selectivity of terahertz gas-phase
  spectroscopy,'' \emph{Analytical Chemistry}, vol.~87, no.~21, pp.
  10\,679--10\,683, Oct. 2015.

\bibitem{Swearer2018}
D.~F. Swearer, S.~Gottheim, J.~G. Simmons, D.~J. Phillips, M.~J. Kale, M.~J.
  McClain, P.~Christopher, N.~J. Halas, and H.~O. Everitt, ``Monitoring
  chemical reactions with terahertz rotational spectroscopy,'' \emph{{ACS}
  Photonics}, vol.~5, no.~8, pp. 3097--3106, May 2018.

\bibitem{Tayvah2021}
U.~Tayvah, J.~A. Spies, J.~Neu, and C.~A. Schmuttenmaer, ``Nelly: A
  user-friendly and open-source implementation of tree-based complex refractive
  index analysis for terahertz spectroscopy,'' \emph{Analytical Chemistry},
  vol.~93, no.~32, pp. 11\,243--11\,250, Aug. 2021.

\bibitem{Ro/nne1997}
C.~Ro/nne, L.~Thrane, P.-O. {\AA}strand, A.~Wallqvist, K.~V. Mikkelsen, and
  S.~R. Keiding, ``Investigation of the temperature dependence of dielectric
  relaxation in liquid water by {THz} reflection spectroscopy and molecular
  dynamics simulation,'' \emph{The Journal of Chemical Physics}, vol. 107,
  no.~14, pp. 5319--5331, Oct. 1997.

\bibitem{Yada2008}
H.~Yada, M.~Nagai, and K.~Tanaka, ``Origin of the fast relaxation component of
  water and heavy water revealed by terahertz time-domain attenuated total
  reflection spectroscopy,'' \emph{Chemical Physics Letters}, vol. 464, no.
  4-6, pp. 166--170, Oct. 2008.

\bibitem{Liebe1991}
H.~J. Liebe, G.~A. Hufford, and T.~Manabe, ``A model for the complex
  permittivity of water at frequencies below 1 {THz},'' \emph{International
  Journal of Infrared and Millimeter Waves}, vol.~12, no.~7, pp. 659--675, Jul.
  1991.

\bibitem{Moeller2009}
U.~M{\o}ller, D.~G. Cooke, K.~Tanaka, and P.~U. Jepsen, ``Terahertz reflection
  spectroscopy of debye relaxation in polar liquids [invited],'' \emph{Journal
  of the Optical Society of America B}, vol.~26, no.~9, p. A113, Aug. 2009.

\bibitem{Fitzgerald2014}
V.~P.~W. Anthony J.~Fitzgerald, Emma Pickwell-MacPherson, ``Use of finite
  difference time domain simulations and debye theory for modelling the
  terahertz reflection response of normal and tumour breast tissue,''
  \emph{PLoS ONE 9(7): e99291}, 2014.

\bibitem{Kindt1996}
J.~T. Kindt and C.~A. Schmuttenmaer, ``Far-infrared dielectric properties of
  polar liquids probed by femtosecond terahertz pulse spectroscopy,'' \emph{J.
  Phys. Chem.}, pp. 10\,373--10\,379, 1996.

\bibitem{Bernier2016}
M.~Bernier, F.~Garet, J.-L. Coutaz, H.~Minamide, and A.~Sato, ``Accurate
  characterization of resonant samples in the terahertz regime through a
  technique combining time-domain spectroscopy and kramers{\textendash}kronig
  analysis,'' \emph{{IEEE} Transactions on Terahertz Science and Technology},
  vol.~6, no.~3, pp. 442--450, May 2016.

\bibitem{strutz2011}
T.~Strutz, \emph{Data fitting and uncertainty: A practical introduction to
  weighted least squares and beyond}.\hskip 1em plus 0.5em minus 0.4em\relax
  Springer, 2011, vol.~1.

\bibitem{Markelz2022}
A.~G. Markelz and D.~M. Mittleman, ``Perspective on terahertz applications in
  bioscience and biotechnology,'' \emph{{ACS} Photonics}, vol.~9, no.~4, pp.
  1117--1126, Apr. 2022.

\bibitem{Peretti2019}
R.~Peretti, S.~Mitryukovskiy, K.~Froberger, M.~A. Mebarki, S.~Eliet,
  M.~Vanwolleghem, and J.-F. Lampin, ``{THz}-{TDS} time-trace analysis for the
  extraction of material and metamaterial parameters,'' \emph{{IEEE}
  Transactions on Terahertz Science and Technology}, vol.~9, no.~2, pp.
  136--149, Mar. 2019.

\bibitem{Mohtashemi2021}
L.~Mohtashemi, P.~Westlund, D.~G. Sahota, G.~B. Lea, I.~Bushfield, P.~Mousavi,
  and J.~S. Dodge, ``Maximum-likelihood parameter estimation in terahertz
  time-domain spectroscopy,'' \emph{Optics Express}, vol.~29, no.~4, p. 4912,
  Feb. 2021.

\bibitem{Dodge2020}
J.~S. Dodge, L.~Mohtashemi, P.~Westlund, P.~Mousavi, and D.~G. Sahota, ``A
  maximum-likelihood analysis framework for terahertz time-domain
  spectroscopy,'' in \emph{2020 45th International Conference on Infrared,
  Millimeter, and Terahertz Waves ({IRMMW}-{THz})}.\hskip 1em plus 0.5em minus
  0.4em\relax {IEEE}, Nov. 2020.

\bibitem{menlo}
MenloSystems, ``Data sheet of teralyzer thz-tds data evaluation software,''
  2017.

\bibitem{Tuzlukov2018}
V.~Tuzlukov, \emph{Signal Processing Noise}, {CRC Press}, Ed., 2018.

\bibitem{peiponen2012terahertz}
K.-E. Peiponen, A.~Zeitler, and M.~Kuwata-Gonokami, \emph{Terahertz
  spectroscopy and imaging}.\hskip 1em plus 0.5em minus 0.4em\relax Springer,
  2012, vol. 171.

\bibitem{mitryukovskiy2019shining}
S.~Mitryukovskiy, M.~Lavancier, F.~Braud, G.~Roos, T.~Hannotte, E.~Dubois,
  J.-F. Lampin, and R.~Peretti, ``Shining the light to terahertz spectroscopy
  of nl-volume biological samples,'' in \emph{CLEO: Applications and
  Technology}.\hskip 1em plus 0.5em minus 0.4em\relax Optical Society of
  America, 2019, pp. ATu3K--6.

\bibitem{qiao2014identification}
L.~Qiao, Y.~Wang, Z.~Zhao, and Z.~Chen, ``Identification and quantitative
  analysis of chemical compounds based on multiscale linear fitting of
  terahertz spectra,'' \emph{Optical Engineering, International Society for
  Optics and Photonics}, vol.~53, no.~7, p. 074102, 2014.

\bibitem{Mitryukovskiy2021}
S.~Mitryukovskiy, D.~E.~P. Vanpoucke, Y.~Bai, T.~Hannotte, M.~Lavancier,
  D.~Hourlier, G.~Roos, and R.~Peretti, ``On the influence of water on thz
  vibrational spectral features of molecular crystals,'' \emph{PCCP (under
  submission)}, 2021.

\bibitem{banks2021necessity}
P.~A. Banks, L.~Burgess, and M.~T. Ruggiero, ``The necessity of periodic
  boundary conditions for the accurate calculation of crystalline terahertz
  spectra,'' \emph{Physical Chemistry Chemical Physics, Royal Society of
  Chemistry}, 2021.

\bibitem{Lavancier2021}
M.~Lavancier, ``Heuristic approach to take up the challenge of terahertz
  time-domain spectroscopy for biology,'' 2021.

\bibitem{Federici2005}
J.~F. Federici, B.~Schulkin, F.~Huang, D.~Gary, R.~Barat, F.~Oliveira, and
  D.~Zimdars, ``Thz imaging and sensing for security applications—explosives,
  weapons and drugs,'' \emph{Semiconductor Science and Technology}, vol.~20,
  no.~7, p. S266–S280, Jun. 2005.

\bibitem{Taday2003}
P.~F. Taday, ``Applications of terahertz spectroscopy to pharmaceutical
  sciences,'' vol. 362, no. 1815, pp. 351--364.

\bibitem{Lee2023}
J.~Lee, C.~K. Leung, M.~Ma, J.~Ward-Berry, S.~Santitewagun, and J.~A. Zeitler,
  ``The dotthz project: A standard data format for terahertz time-domain
  data,'' \emph{Journal of Infrared, Millimeter, and Terahertz Waves}, vol.~44,
  no.~11, pp. 795--813, 2023.

\end{thebibliography}
	\vfill

\end{document}